# Improved Interfacial and Electrical Properties of GaSb Metal Oxide Semiconductor Devices Passivated with Acidic $(NH_4)_2S$ Solution

Lianfeng Zhao, Zhen Tan, Jing Wang, and Jun Xu[*]

Tsinghua National Laboratory for Information Science and Technology, Institute of Microelectronics, Tsinghua University, Beijing 100084, China

[*]Email: junxu@tsinghua.edu.cn

## Abstract

Surface passivation with acidic $(NH_4)_2S$ solution is shown to be effective in improving the interfacial and electrical properties of $HfO_2$/GaSb metal oxide semiconductor devices. Compared with control samples, those treated with acidic $(NH_4)_2S$ solution showed great improvements in frequency dispersion, gate leakage current and interface trap density. These improvements were attributed to the acidic $(NH_4)_2S$ solution enhancing passivation of the substrates, which was analyzed from the perspective of chemical mechanism and confirmed by X-ray photoelectron spectroscopy and high-resolution cross-sectional transmission electron microscopy.

To extend Moore's law of complementary metal oxide semiconductor (CMOS) scaling, high-mobility III-V semiconductor materials, such as GaAs and InGaAs, have been actively investigated as alternative channel materials to Si.[1] Although these III-V materials possess high electron mobility,[2] their hole mobility is low.[3] Consequently, GaSb has attracted considerable attention as a channel material for p-type MOS devices because of its high hole mobility.[4] However, the interface trap density ($D_{it}$) of GaSb MOS devices is still too high for their wide application.[5] Many methods have been proposed to improve the interfacial properties of III-V MOS devices, such as fluorine passivation,[6] post-deposition annealing,[7] plasma treatment,[8] insertion of an interfacial layer,[9] and the use of novel dielectrics.[10] We previously showed that a lower $D_{it}$ can be obtained for devices passivated with neutralized $(NH_4)_2S$ solution.[11] To obtain a more comprehensive understanding of the $(NH_4)_2S$ passivation technique, here we investigate the effects of the surface passivation treatment with acidic $(NH_4)_2S$ solution on GaSb MOS capacitors (MOSCAPs). We found that treatment with acidic $(NH_4)_2S$ solution is superior to that with 20% $(NH_4)_2S$ solution, which has a pH of ~11 and is hereafter referred to as basic $(NH_4)_2S$ solution. Compared with control samples, those treated with acidic $(NH_4)_2S$ solution showed great improvements in frequency dispersion, gate leakage current and interface trap density.

Te-doped (100)-oriented n-type GaSb wafers with a doping concentration of ~$10^{17}$ cm$^{-3}$ were used as substrates. The substrates were first degreased by sequential immersion for 5 min each in acetone, ethanol, and isopropanol, and then cleaned with 9% HCl aqueous solution for 1 min. Sulfur passivation treatments were performed for 15 min using basic and acidic $(NH_4)_2S$ solutions for different groups of samples. An $HfO_2$ dielectric layer (~5.5 nm) was then deposited on each sample by atomic layer deposition (ALD) at 200 °C using tetrakis(ethylmethylamino)hafnium (TEMAH) and water as precursors. Finally, Al was evaporated and patterned to form MOSCAPs. Back metal contacts of Ti/Au were also deposited. Acidic $(NH_4)_2S$ solution was prepared by adding 37% HCl aqueous solution to 20% $(NH_4)_2S$ aqueous solution until the pH of the acidic $(NH_4)_2S$ solution was 5.5–6.0. It should be noted that $(NH_4)_2S$ solution was not stable at pH lower than 5.5. High-resolution X-ray photoelectron spectroscopy (XPS) was performed to investigate the chemical states of the surface passivated with basic and acidic $(NH_4)_2S$ solution. The $HfO_2$/GaSb MOS stacks were physically characterized by high-resolution transmission electron microscopy (HRTEM). Capacitance-voltage (C-V), conductance-voltage (G-V) and gate leakage current-voltage (J-V) characteristics were recorded using an Agilent B1500A semiconductor device analyzer and a Cascade Summit 11000 AP probe system.

Fig. 1(a) and (b) show XPS Ga 3d spectra of GaSb surfaces passivated with basic and acidic $(NH_4)_2S$ solutions, respectively. The Ga-O peak at 20.3 eV is very intense for the sample treated with basic $(NH_4)_2S$ solution,[12] and reduces in intensity considerably for the sample treated with acidic $(NH_4)_2S$ solution. This indicates that the acidic $(NH_4)_2S$ solution can effectively remove and prevent the formation of Ga-O bonds. This is because in the acidic solution, the chemical reaction $GaO_x + H^+ + S^{2-} \rightarrow GaS_x + H_2O$ take place, preventing surface oxidation and promoting the sulfur passivation process. This reaction cannot occur in basic $(NH_4)_2S$ solution because the concentration of protons is much lower. XPS Sb $3d_{3/2}$ spectra of the GaSb surfaces passivated with basic and acidic $(NH_4)_2S$ solutions are presented in Fig. 2(a) and (b), respectively. The Sb-O bonds at binding energies of 539.6 eV ($Sb_2O_3$) and 540.2 eV ($Sb_2O_5$) do not change obviously for samples treated with acidic $(NH_4)_2S$ solutions compared with those treated with basic $(NH_4)_2S$ solution.[13] This is because the residue oxides after basic $(NH_4)_2S$ solution treatment are mainly gallium oxide, and the intensities of the signals from Sb-O bonds are very weak. This is consistent with the fact that the Sb-O phase is not thermodynamically stable on the GaSb surface and the gallium oxide component of the native oxides on the GaSb surface is much higher than that of the antimony oxides.[14] The intensity of the Sb-S peak shown at 539.1 eV is identical for each sample.[13] The similarity of the XPS Sb $3d_{3/2}$ spectra for samples treated with basic and acidic $(NH_4)_2S$ solutions indicates that basic and acidic $(NH_4)_2S$ solutions have similar effect on Sb related passivation.

The thinner surface oxide layer present in samples treated with acidic $(NH_4)_2S$ solution than those treated with basic $(NH_4)_2S$ solution was also confirmed by HRTEM, as shown in Fig. 3. An oxide layer, which is about 1.8 nm thick, can be clearly observed between the $HfO_2$ dielectric and GaSb substrate for the sample treated with basic $(NH_4)_2S$ solution (see Fig. 3(a)). In contrast, this surface oxide layer is ~0.5 nm thick for the sample treated with acidic $(NH_4)_2S$ solution, as illustrated in Fig. 3(b). The thinner surface oxide layer for the sample treated with the acidic $(NH_4)_2S$ solution indicates that the acidic $(NH_4)_2S$ solution has a more favorable effect on surface passivation than the basic $(NH_4)_2S$ solution. According to the XPS results presented in Figs. 1 and 2, the main component of this surface oxide layer is gallium oxide. It should be noted that an interfacial layer also formed between the Al electrode and $HfO_2$ layer, which might be caused by the reaction of Al and $HfO_2$, considering the larger negative Gibbs free energy to form

Al$_2$O$_3$ than HfO$_2$.[15] This interfacial layer will lead to a larger equivalent oxide thickness (EOT).

Fig. 4 shows the multi-frequency C-V characteristics of GaSb MOSCAPs passivated with basic and acidic (NH$_4$)$_2$S solutions. The frequency dispersion in the accumulation region is reduced considerably by passivation with acidic (NH$_4$)$_2$S solution instead of the basic solution, which indicates that the interfacial properties are improved using the acidic solution.[16] The EOT for the samples treated with basic and acidic (NH$_4$)$_2$S solutions is 3.46 and 2.60 nm, respectively, which is higher than the expected EOT for the ~5 nm HfO$_2$ layer. The higher EOT should be caused by the interfacial layer formed between the Al electrode and HfO$_2$ layer (see Fig. 3). Compared with the EOT of samples treated with basic (NH$_4$)$_2$S solution, the lower EOT of samples treated with acidic (NH$_4$)$_2$S solution indicates that the acidic solution can remove the surface oxide layer on the GaSb substrate more effectively than the basic one, which is consistent with the XPS and TEM results. The insets in Fig. 4 show CV hysteresis characteristics of samples treated with basic and acidic (NH$_4$)$_2$S solutions, respectively. A large hysteresis is observed for samples treated with basic (NH$_4$)$_2$S solution, which indicates large border traps.[17] This might be due to the intermixing of the high-k and the interfacial layer.[18] CV hysteresis for samples treated with acidic (NH$_4$)$_2$S solution is reduced significantly, which indicates that border traps are considerably reduced.

The gate leakage current-voltage (*I-V*) characteristics of samples treated with basic and acidic (NH$_4$)$_2$S solutions were examined based on a statistical study; the results are presented in Fig. 5. The inset in Fig. 5 shows box plots of the measured gate leakage current density of different samples at ±4 MV/cm. The small variation of leakage current indicates the good repeatability of the devices. The gate leakage current density at ±4 MV/cm of samples treated with acidic $(NH_4)_2S$ solution is about one order of magnitude lower than that of samples treated with basic $(NH_4)_2S$ solution. This might be caused by the reduction of trap-assisted current owing to the reduction of border traps and interface traps.[18]

Fig. 6(a) shows typical measured parallel $G_p/\omega$ *versus* frequency curves for different gate biases of the MOSCAPs passivated with acidic $(NH_4)_2S$ solution. The observed peak shift indicates the efficiency of the surface Fermi level movement over the energy gap. $D_{it}$ distribution in the upper half of the band gap was determined using the conductance method,[19] as plotted in Fig. 6(b). Low temperature (77 K) measurement was performed to investigate the $D_{it}$ distribution over a wider range of the band gap. The extracted $D_{it}$ distribution is consistent with previously reported results.[5] Compared with the samples treated with basic $(NH_4)_2S$ solution, $D_{it}$ is reduced by ~53% for the samples treated with acidic $(NH_4)_2S$ solution, which is consistent with the smaller frequency dispersion of the samples treated with acidic solution (see Fig. 4).

In summary, sulfur passivation with acidic $(NH_4)_2S$ solutions was found to be a promising method to improve the properties of GaSb MOS devices. Frequency dispersion, gate leakage current and $D_{it}$ are all reduced for samples treated with acidic $(NH_4)_2S$ solution compared with those treated with basic $(NH_4)_2S$ solution. These improvements can be explained by acidic $(NH_4)_2S$ solution reducing $GaO_x$ interfacial layer and enhancing sulfur passivation on the surface of the GaSb substrates compared with basic $(NH_4)_2S$ solution.

**Acknowledgments:** This work was supported in part by the State Key Development Program for Basic Research of China (No. 2011CBA00602) and by the National Science and Technology

**Figure Captions**

FIG. 1. XPS Ga 3d spectra of GaSb MOSCAPs passivated with (a) basic and (b) acidic $(NH_4)_2S$ solutions.

FIG. 2. XPS Sb $3d_{3/2}$ spectra of GaSb MOSCAPs passivated with (a) basic and (b) acidic $(NH_4)_2S$ solutions.

FIG. 3. HRTEM images of GaSb MOSCAPs passivated with (a) basic and (b) acidic $(NH_4)_2S$ solutions.

FIG. 4. Multi-frequency C-V characteristics of GaSb MOSCAPs passivated with (a) basic and (b) acidic $(NH_4)_2S$ solutions. The insets show CV hysteresis characteristics of the samples.

FIG. 5. Comparison of gate leakage current characteristics of MOSCAPs passivated with basic and acidic $(NH_4)_2S$ solutions. The inset shows box plots of the gate leakage current density at ±4 MV/cm.

FIG. 6. (a) Typical measured parallel $G_p/\omega$ *versus* frequency curves for different gate biases of the MOSCAPs passivated with acidic $(NH_4)_2S$ solution. (b) $D_{it}$ distributions of MOSCAPs passivated with basic and acidic $(NH_4)_2S$ solutions.

Figures

Fig.1

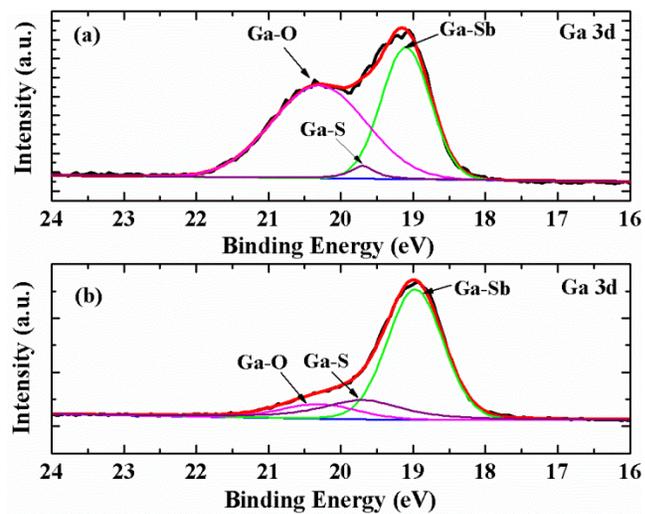

Figures (continued)

Fig.2

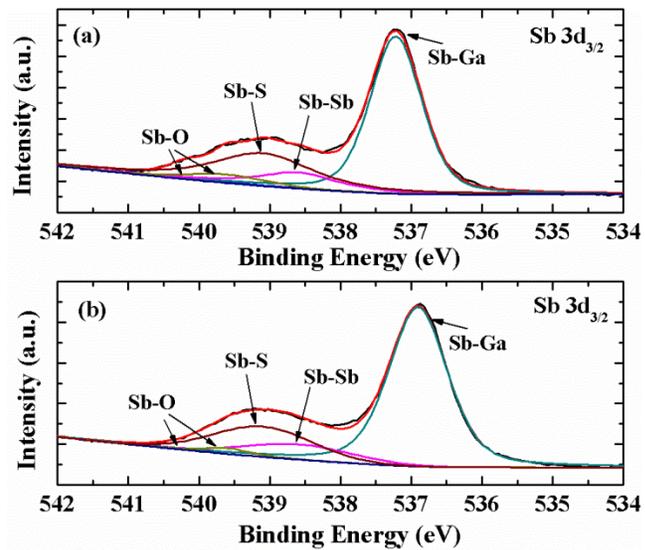

Figures (continued)

Fig.3

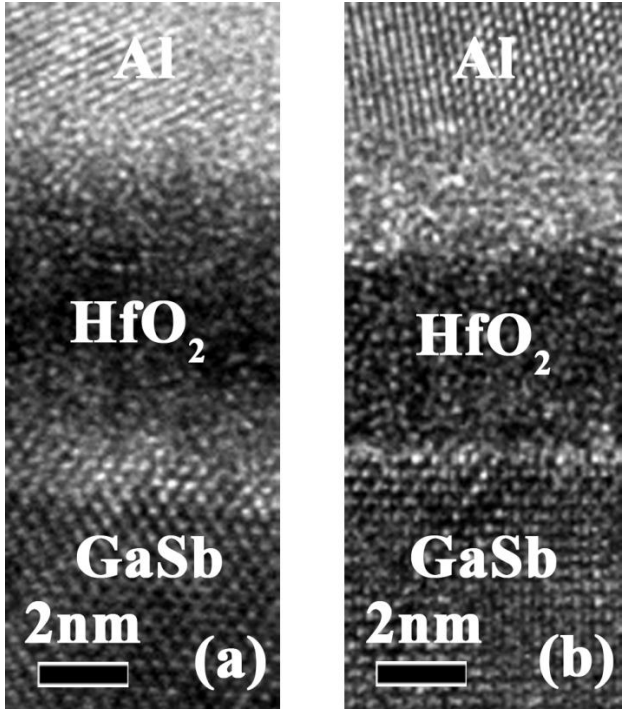

Figures (continued)

Fig.4

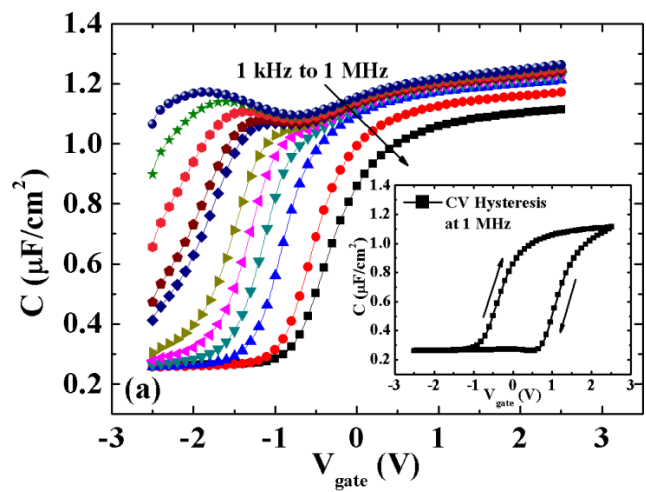

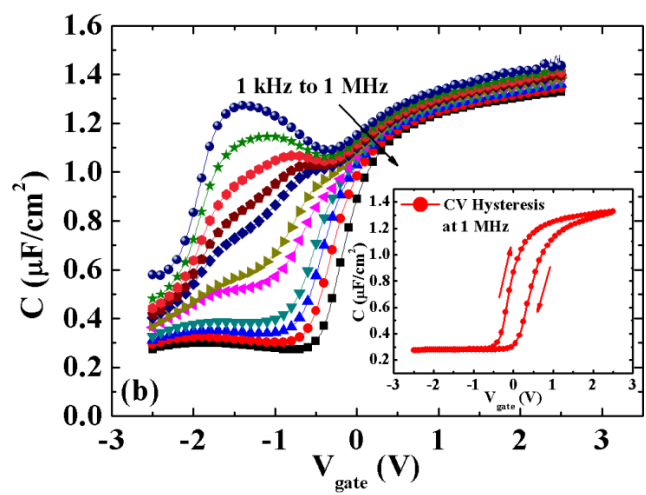

Figures (continued)

Fig.5

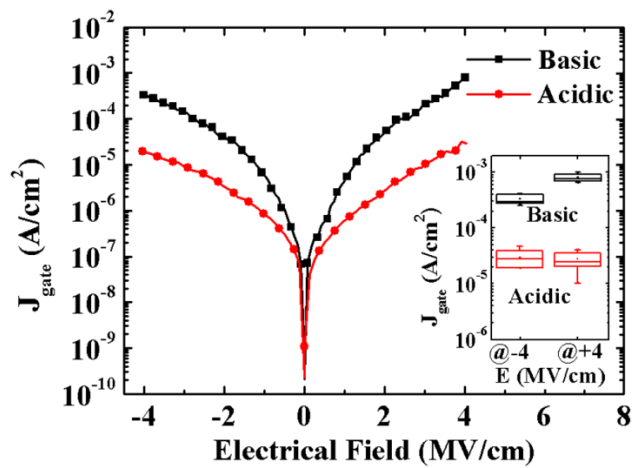

Figures (continued)

Fig.6

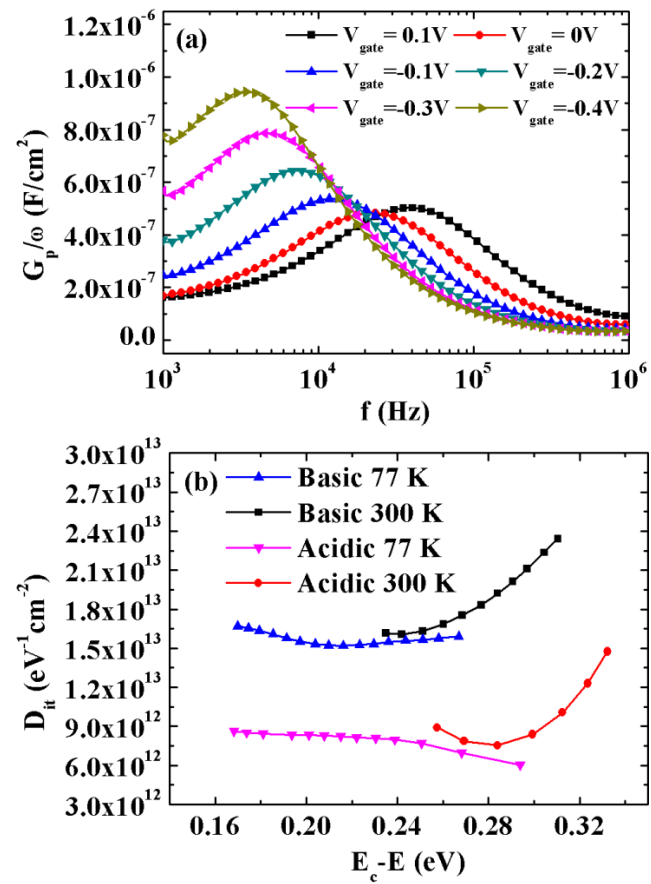